\documentclass[prb,aps, twocolumn, superscriptaddress]{revtex4-1}
\usepackage{amsmath,amssymb}
\usepackage{graphicx}
\usepackage{epsfig, color}
\usepackage{wasysym}
\usepackage{amsfonts}
\usepackage{bm}
\usepackage{enumerate}
\usepackage{color}
\usepackage[resetlabels]{multibib}
\usepackage{epstopdf}
\usepackage{latexsym}
\usepackage[breaklinks,colorlinks = true,linkcolor = red,urlcolor=cyan,citecolor=red]{hyperref}
\usepackage[caption=false]{subfig}
\usepackage{float}

\newcommand{\bea}{\begin{eqnarray}}
\newcommand{\eea}{\end{eqnarray}}
\newcommand{\be}{\begin{eqnarray}}
\newcommand{\ee}{\end{eqnarray}}
\newcommand{\bw}{\begin{widetext}}
\newcommand{\ew}{\end{widetext}}

\newcommand{\tmod}[1]{\,\text{mod $#1$}}

\newcommand{\expval}[1]{ \langle #1 \rangle}

\newcommand{\matel}[3]{\langle #1 | #2 | #3 \rangle}

\newcommand{\pfC}[0]{\alpha^{\vphantom{\dagger}}}
\newcommand{\pfCconj}[0]{\alpha^{\dagger}}

\usepackage[normalem]{ulem}

\def\ket#1{{|#1\rangle}}
\def\bra#1{{\langle #1 |}}

\begin{document}
\title{Localization-protected order  in  spin chains with  non-Abelian discrete symmetries}
\author{Aaron J. Friedman}
\affiliation{The Rudolf Peierls Centre for Theoretical Physics, University of Oxford, Oxford OX1 3NP, UK}
\affiliation{Department of Physics and Astronomy, University of California, Irvine, CA 92697, USA}
\author {Romain Vasseur}
\affiliation{Department of Physics, University of Massachusetts, Amherst, MA 01003, USA}
\affiliation{Department of Physics, University of California, Berkeley, CA 94720, USA}
\affiliation{Materials Science Division, Lawrence Berkeley National Laboratories, Berkeley, CA 94720}
\author{Andrew C. Potter}
\affiliation{Department of Physics, University of Texas at Austin, TX 78712, USA}
\author{S. A. Parameswaran}
\affiliation{The Rudolf Peierls Centre for Theoretical Physics, University of Oxford, Oxford OX1 3NP, UK}
\affiliation{Department of Physics and Astronomy, University of California, Irvine, CA 92697, USA}

\date{\today}
\begin{abstract}

{We  study the non-equilibrium phase structure of the three-state random quantum Potts model in one dimension. This spin chain is characterized by a non-Abelian $D_3$ symmetry recently argued to be incompatible with the existence of a symmetry-preserving many-body localized (MBL) phase. Using exact diagonalization and a finite-size scaling analysis, we find that the model supports two distinct broken-symmetry MBL phases at strong disorder that either break the ${\mathbb{Z}_3}$ clock symmetry or a ${\mathbb{Z}_2}$ chiral symmetry. In a dual formulation, our results indicate the existence of a stable finite-temperature topological phase with MBL-protected parafermionic end zero modes. While we find a thermal symmetry-preserving regime for weak disorder,  scaling analysis at strong disorder points to an infinite-randomness critical point between two distinct broken-symmetry MBL phases.}

\end{abstract}
\maketitle

\section{Introduction} Little is known about the generic properties and possible phases of quantum systems out of equilibrium, where even a basic understanding at the level of Landau theory has remained elusive. Many-body systems can reach an effective equilibrium state under their own dynamics even in isolation, a process encoded in the properties of individual eigenstates via a set of criteria collectively termed the {eigenstate thermalization hypothesis} (ETH) ~\cite{PhysRevA.43.2046, PhysRevE.50.888}. A distinct class of {many-body localized} (MBL) systems~\cite{BAA,ARCMP}, usually with quenched randomness,  violate these criteria and  cannot self-thermalize. As they need not satisfy the stringent requirements imposed by ETH, even highly excited eigenstates of MBL systems can exhibit properties usually associated with quantum ground states~\cite{HuseMBLQuantumOrder,BauerNayak}. This permits the classification of out-of-equilibrium  MBL systems into distinct {\it eigenstate phases} separated by {\it eigenstate phase transitions}~\cite{EPTReview}, echoing the classification of ground states and critical points in equilibrium systems, and providing a window into non-equilibrium quantum order.

As ETH systems conform to the expectations of equilibrium statistical mechanics, at infinite effective temperature ($T\rightarrow \infty)$ they exhibit the  eigenstate phase structure of a thermal paramagnet, 
with trivial spatial and temporal correlations. In contrast, MBL systems can exhibit richer behavior --- for instance, an MBL Ising model can exhibit both
broken-symmetry (spin glass) and paramagnetic phases even for $T\rightarrow \infty$~\cite{HuseMBLQuantumOrder,PhysRevLett.112.217204,KjallIsing,PekkerRSRGX}. 
In this regime --- that is dominated by properties of highly-excited eigenstates --- a particularly sharp distinction emerges for {\it non-Abelian} symmetries. While ETH systems with such symmetries are again thermal paramagnets as $T\rightarrow\infty$, a fully MBL phase is inconsistent with non-Abelian symmetry~\cite{PhysRevB.94.224206}.  This indicates that the onset of full MBL must coincide with breaking of the symmetry, and that any symmetry-preserving phase is either (i) thermal, in the sense of ETH; or (ii) an athermal `quantum critical glass'~\cite{QCGPRL} that does not admit the local tensor product description of a fully MBL phase.
In recent work~\cite{XXZPaper}, we illustrated 
the former scenario in fermion chains with $U(1)\rtimes {\mathbb{Z}_2}$ symmetry, whose non-Abelian semi-direct product structure (denoted `$\rtimes$') reflects the nontrivial action of a ${\mathbb{Z}_2}$ particle-hole symmetry (PHS) on a conserved $U(1)$ charge. As disorder is increased, PHS is spontaneously broken in highly excited eigenstates whenever they are fully MBL. 
This rules out the possibility of using PHS in conjunction with MBL to stabilize a  topological phase --- corresponding to class AIII in the usual taxonomy~\cite{PerioTableSchnyder, PerioTableKitaev} of symmetry-protected topological (SPT) phases --- as $T\rightarrow \infty$. Similar considerations~\cite{PhysRevB.94.224206} limit MBL-protected SPT phases~\cite{PhysRevB.89.144201,BahriMBLSPT,CenkeMBLSPT,ACPMBLSPT}, and rule out many phases that host non-Abelian anyons.

Some 1D  topological phases host non-Abelian parafermionic edge modes {\it independent} of any protecting symmetries~\cite{AliceaFendleyReview}, and are therefore apparently more amenable to localization protection. However, such 1D parafermionic chains are still inextricably linked to symmetry, although indirectly: they are related, via the Fradkin-Kadanoff mapping~\cite{FRADKIN19801}, to $\mathbb{Z}_n$ quantum clock models, with the topological (trivial) phases of the parafermions corresponding to ordered (disordered) phases of the spins~\cite{AliceaFendleyReview}. The question of MBL protection of parafermion edge modes then turns on the interplay of localization with the symmetry of the spin chain.  Though $\mathbb{Z}_n$ itself is Abelian, for special {\it achiral} parameter values, ${\mathbb{Z}_n}$ clock models with $n\geq 3$ acquire an additional ${\mathbb{Z}_2}$ reflection symmetry, enhancing the global symmetry to that of the non-Abelian dihedral group $D_n \cong {\mathbb{Z}_n} \rtimes {\mathbb{Z}_2}$. For $n=3$, since $D_3 = S_3$, this is equivalent to the 3-state quantum Potts model. Our focus here is on understanding how the non-Abelian  symmetry of this model influences its non-equilibrium phase structure.

This question is interesting for several reasons, explored in the remainder of this paper. First, while superficially similar to the previously-studied $U(1)\rtimes {\mathbb{Z}_2}$ example, the $D_3$ model has a richer phase diagram --- intuitively, $U(1)$ symmetry constrains possible phases more strongly than  $D_3$ symmetry.  Second, since breaking $U(1)$ symmetry is impossible even with MBL in $d=1$, the {\it only} possibility for an MBL phase in a $U(1)\rtimes {\mathbb{Z}_2}$ system breaks the ${\mathbb{Z}_2}$ symmetry; in the clock model, the ${\mathbb{Z}_3}$ subgroup of $D_3$ may also be broken. This means that at strong disorder it may be possible to tune between distinct MBL phases via an unusual critical point.  Third, there is the intriguing possibility of an athermal, symmetry-preserving quantum critical glass phase [case (ii) discussed above]. Finally, {to return to our original motivation}, ${\mathbb{Z}_3}$ breaking in highly excited eigenstates translates via duality~\cite{Parafendlions} to the existence of a non-equilibrium topological phase that hosts edge parafermion modes relevant to fault-tolerant quantum computing. 

Before proceeding, we mention that  related work by Prakash \emph{et al.}\cite{SUNYPotts}, also considered the role of non-Abelian symmetries in the excited states of a somewhat distinct model. Our results are consistent with those, when they overlap, though their emphasis was not on the nature of the critical behavior between MBL phases. Their work also speculates as to the existence of a QCG phase, a question on which we remain for the moment agnostic.

\section{Model and Symmetries} We now turn to the ${\mathbb{Z}_3}$  quantum clock model, described by the Hamiltonian {
\be \label{eq:Ham} H =  -\sum_{j=1}^{L-1}J_j  e^{i \phi_j} {\hat{\sigma}}^{\dagger}_j {\hat{\sigma}}^{\vphantom{\dagger}}_{j+1} -  \sum_{j=1}^Lf_j e^{i\theta_j} {\hat{\tau}}^{\vphantom{\dagger}}_j    + \text{h.c.},\ee}%\right),\ee
where $J_j$, $f_j$, $\theta_j$, $\phi_j$ are all real, and discussed below. The operators commute on different sites and satisfy 
\be
{\hat{\sigma}}_j^3 = {\hat{\tau}}_j^3=1, ~ {\hat{\sigma}}_j {\hat{\tau}}_j = \omega {\hat{\tau}}_j {\hat{\sigma}}_j,
\ee
on a single site, where $\omega = e^{\frac{2 \pi i}3}$. 

In the eigenbasis of the {\it weight} operator, ${\hat{\sigma}}$, defined by: ${\hat{\sigma}} \ket{m} = \omega^{m} \ket{m}$, ${\hat{\tau}}$ is a {\it shift} operator:  ${\hat{\tau}} \ket{m} = \ket{m+1}$, with ket labels taken modulo $3$ henceforth. The conjugate $\hat{\tau}$-eigenbasis $\ket{q}$ interchanges these roles: ${\hat{\tau}} \ket{q}= \omega^{q}\ket{q}$, ${\hat{\sigma}} \ket{q} = \ket{q-1}$. Viewed as a 3-state quantum rotor, $-i \ln \hat{\sigma}$ represents the angle of the rotor, and the $\hat{\tau}$ measures its angular momentum (modulo 3).

For generic parameter choices, \eqref{eq:Ham} has a global ${\mathbb{Z}_3}$ rotation symmetry generated 
by ${\hat{\mathcal{Q}}} = \prod_j {\hat{\tau}}_j$, with ${\hat{\mathcal{Q}}}^3=1$. For  $\phi_j,~\theta_j \equiv 0 \tmod{\pi/3}$, there is also a {$\mathbb{Z}_2$ mirror symmetry:} ${\hat{\mathcal{X}}} \equiv \prod_j {\hat{\mathcal{X}}}_j$, where ${\hat{\mathcal{X}}}_j$ exchanges the $\ket{1}$, $\ket{2}$ eigenstates of \emph{either} $\hat{\sigma}_j$, $\hat{\tau}_j$, and ${\hat{\mathcal{X}}}^2 =1$. Together, ${\hat{\mathcal{Q}}}, {\hat{\mathcal{X}}}$ generate the group $D_3 = S_3 \cong {\mathbb{Z}_3}\rtimes {\mathbb{Z}_2}$, where the semidirect product structure reflects the fact that ${\hat{\mathcal{Q}}}$,  ${\hat{\mathcal{X}}}$ do not commute. Consequently, in the $\sigma_j$ basis there are two additional ${\mathbb{Z}_2}$ symmetries ${\hat{\mathcal{X}}\hat{\mathcal{Q}}}$, ${\hat{\mathcal{X}}\hat{\mathcal{Q}}}^2$, which respectively leave $\ket{1}$, $\ket{2}$ invariant, while exchanging the other states.
{When they are viewed as $3$-state quantum rotors in the $xy$-plane, $\hat{\mathcal{Q}}$ is a $2\pi/3$ rotation, and $\hat{\mathcal{X}}$ is a mirror reflection of the rotor about one of its $3$ directions (which also inverts the angular momentum, $\hat{\tau}$).}

In the clean limit of the $\mathbb{Z}_3$ Hamiltonian \eqref{eq:Ham}, the ground state has an ordered phase that spontaneously breaks $\hat{\mathcal{Q}}$ for $J \gg f$, and a disordered phase for $J\ll f$. These correspond respectively to parafermionic phases with and without edge zero modes~\cite{AliceaFendleyReview}. The chiral couplings explicitly break $\mathbb{Z}_2$, and the ground states in both limits break two of the ${\mathbb{Z}_2}$ reflection symmetries ${\hat{\mathcal{X}}},{\hat{\mathcal{X}}\hat{\mathcal{Q}}}, {\hat{\mathcal{X}}\hat{\mathcal{Q}}}^2$.  This model also possesses a sequence of incommensurate phases~\cite{OstlundIncommensurate,HuseIncommensurate}. Extensive recent work~\cite{MotrukPara,BondesanQuellaParafermion,MengPara,BernevigPara,z3ChiralPD} has focused on the clean case; 
we shall  instead study the situation when the couplings are {disordered}, i.e. $J_j$ and $f_j$ on each site are i.i.d. random variables, while keeping $\phi_j, \theta_j \equiv 0 \tmod{\pi/3}$, so that the symmetry group is $D_3$. We wish to understand whether highly excited eigenstates of \eqref{eq:Ham} satisfy ETH, or are instead MBL; and if the latter, whether and how they break the non-Abelian $D_3$ symmetry. We note that previous analysis~\cite{JermynZMStab} of {low-energy excited-states} of clock models focused on edge zero modes in clean systems, a setting quite distinct from the {non-equilibrium} disordered case studied here. For the non-Abelian XXZ chain~\cite{XXZPaper}, the excited-state real-space renormalization group~\cite{PekkerRSRGX} (RSRG-X) provides useful insights; here, the reduced $D_3$ symmetry complicates matters, as explained in Appendix~\ref{subsec:s3rg}. Since RSRG-X is inconclusive, we turn instead to a numerical analysis, which we now describe.

\section{Numerics} We investigate the random $D_3$ chain via numerical exact diagonalization of \eqref{eq:Ham}. Since we ahve three states per site, only a discrete global symmetry, and study highly excited eigenstates, we are limited to systems of length $L\leq 10$. Therefore, we must perform  finite-size scaling analysis of our data in order to extract the phase diagram and conjectured critical behavior.  It is convenient to parametrize couplings as $J_j = {\lambda_j}\frac{(1 + \delta)}{2}$ and $f_j = {\lambda_j'}\frac{(1 - \delta)}{2}$; with this choice, $\delta=\pm1$ are trivial limits corresponding to idealized fixed-point Hamiltonians for the ordered (spin glass) and paramagnetic phases respectively. (In the parafermionic language, $\delta$ is the dimerization, i.e. the bias in strength between odd and even couplings.) The random coefficients $\lambda,\lambda'$ are drawn from the distribution $P(\lambda) = \frac{1}{W} \lambda^{1-1/W}$, where disorder is stronger for larger $W$, with $W=1.0$ equivalent to a uniform distribution on $(0,1]$. For each realization of disorder we use the shift-invert {method~\cite{Luitz}} to obtain $\mathcal{N}=50$ eigenstates from approximately the middle of the many-body spectrum, and average our data over $10^3 -10^4$ such realizations. We use open boundary conditions, and slightly reduce the Hilbert space size by restricting to eigenstates of both ${\hat{\mathcal{Q}}}$ and ${\hat{\mathcal{X}}}$ with eigenvalue unity. 
%%% AARON: the following sentence was added in response to Referee A's comments
Although ${\hat{\mathcal{Q}}}$ and ${\hat{\mathcal{X}}}$  do not commute, states constructed in the $\hat{\tau}$ basis have eigenvalue $\omega^Q$ under ${\hat{\mathcal{Q}}}$, where $Q = \sum_j q_j ~ {\rm mod}~ 3$, and if $Q=0$ for some such state, then the $\mathbb{Z}_2$ partner of this state also has $Q=0$. In this way, we can construct a basis of simultaneous eigenstates of ${\hat{\mathcal{Q}}}$ and ${\hat{\mathcal{X}}}$ from states with ${\hat{\mathcal{Q}}} \ket{\psi} = \ket{\psi}$ by superposing such states with their $\mathbb{Z}_2$ partner.

\subsection{Observables}
To map out the ergodic and localized regions in the transverse field-disorder plane (here parametrized by $\delta, W$ respectively), we numerically measure several indicative quantities.  We study the energy level statistics via the {`$r$-ratio'~\cite{PalHuse}}, defined in terms of gaps $\delta_n = E_n -E_{n-1}$ between successive energy eigenvalues as $r = \min(\delta_n,\delta_{n-1})/\max(\delta_n,\delta_{n-1})$. Once all symmetries have been taken into account, for ETH systems energy level repulsion results in $r\approx 0.53$, characteristic of the Gaussian orthogonal random matrix ensemble, whereas for MBL systems $r \approx 0.38$, reflecting the Poisson statistics when level repulsion is absent~\cite{PhysRevB.75.155111, PalHuse}. 
%%% AARON: the following features changes in response to Referee A's comments
It is crucial that we consider only eigenstates within a given symmetry sector, as pairing between different sectors will artificially suppress $r$~\cite{HuseMBLQuantumOrder, XXZPaper} in broken-symmetry states, and our auxiliary $\mathbb{Z}_2$ order parameter is only valid when evaluated with eigenstates of the $\mathbb{Z}_3$ cycle. We do not show level statistics data for $|\delta|\gtrsim 0.7$ as they show unphysically small $r$-ratios due to  `fragmentation' of the spectrum in the vicinity of perfect dimerization. We also study the scaling of the half-chain entanglement entropy $S_E^{(n)} =-\text{Tr} {\hat{\rho}_n}\ln {\hat{\rho}_n}$, computed in the $n^{\text{th}}$ eigenstate from the reduced density matrix ${\hat{\rho}_n} \equiv \text{Tr}_{i> \lceil L/2\rceil } \ket{n}\bra{n}$.

\subsection{Order parameters}
Glassy breaking of the ${\mathbb{Z}_3}$ symmetry may be diagnosed by an Edwards-Anderson-type order parameter, 
\be\label{eq:s3op}
m_3 = \frac{1}{\mathcal{N}L^2} \sum_{n=1}^{\mathcal{N}}\sum_{i\neq j} \left|\bra{n}{\hat{\sigma}}^{\dagger}_i {\hat{\sigma}}^{\vphantom{\dagger}}_j \ket{n}\right|^2,
\ee
where $n$ labels eigenstates. This is analogous to the order parameter used to analyze MBL phases in the Ising~\cite{KjallIsing} and XXZ~\cite{XXZPaper} chains, and will be non-zero in an eigenstate only if ${\mathbb{Z}_3}$ symmetry is broken. The square average ensures that quenched site-to-site and eigenstate-to-eigenstate variations do not cancel, a standard strategy employed for spin-glass order. Breaking ${\mathbb{Z}_3}$ automatically breaks ${\mathbb{Z}_2}$; we verify this explicitly in our numerics, and will justify analytically in Appendix \ref{subsec:subsidZ2}. We term a phase with ${m_{3}\neq 0}$ a {\it ${\mathbb{Z}_3}$ spin glass}. {In the parafermion language, this $\mathbb{Z}_3$ spin glass corresponds to the topological phase with parafermionic edge states.}

%%%AARON: the following sentence has been changed based on feedback from Referee A
 In addition, there is a ${\mathbb{Z}_2}$ chiral symmetry, $\hat{\mathcal{X}}$, in the transverse-field phase $f\gg J$, in which $\mathbb{Z}_3$ is \emph{preserved}. To detect glassy chiral ordering, we examine
\be
\label{eq:z2op}
m_{\chi} = \frac{1}{\mathcal{N}L^2}  \sum_{n=1}^{\mathcal{N}}\sum_{i\neq j}  \left|\bra{n}{\hat{\mathcal{J}}}_i {\hat{\mathcal{J}}}_j \ket{n}\right|^2,
\ee
where the operator ${\hat{\mathcal{J}}}_j = \frac{1}{i \sqrt{3}} \left( {\hat{\tau}}^{\vphantom{\dagger}}_j - {\hat{\tau}}^{\dagger}_j \right) = \frac{2}{\sqrt{3}} \text{Im}({\hat{\tau}}_j)$ measures the chirality on a single site and anticommutes with ${\hat{\mathcal{X}}} \hat{\mathcal{Q}}^{n}$. We term a phase with $m_{\chi}\neq0$ but $m_3=0$ a {chiral ${\mathbb{Z}_3}$ paramagnet} since it preserves ${\mathbb{Z}_3}$ but breaks the chiral ${\mathbb{Z}_2}$ symmetry. {In the parafermionic representation, the chiral paramagnet corresponds to a topologically trivial MBL phase without edge states.} Observe that $D_3$ symmetry is broken in both the ${\mathbb{Z}_3}$ spin glass and the chiral paramagnet: completely in the former, and down to an Abelian ${\mathbb{Z}_3}$ subgroup in the latter. Note however that $m_\chi =0$ in the ${\mathbb{Z}_3}$ spin glass even though ${\mathbb{Z}_2}$ is broken, as we have defined $m_\chi$ in the dual basis, which we elucidate in Appendix \ref{sec:OPs}.

\subsection{Scaling exponents}
\label{subsec:scaling}
In the strongly disordered limit $W \gg 1$, we conjecture that a direct transition between an MBL spin glass and a chiral MBL paramagnet occurs at the self-dual point $\delta = 0$, and is characterized by ``random singlet'' critical exponents. In particular, we expect this transition to share the universal properties of the $T=0$ disordered Ising chain~\cite{SenthilPotts}
, including a \emph{true} correlation length with mean scaling exponent $\nu \geq 2$, consistent with Harris and Chayes-Chayes-Fisher-Spencer bounds\cite{Chayes}, where $\xi \sim \Delta^{-\nu}$ and $\Delta$ is a disorder-dependent logarithmic measure of distance from criticality~\cite{FisherRSRG1,FisherRSRG2}. 

Additionally, the \emph{mean critical correlations} behave like $\bar{C}_{ij} = \left| \expval{\hat{\sigma}^z_i \hat{\sigma}^z_j} \right| \sim \left| i - j \right|^{-\beta }$. For the random Ising case~\cite{FisherRSRG1,FisherRSRG2}, the exponent $\beta$ has a known value of $\beta_{\rm Ising} = 2-\varphi  $, where $\varphi$ is the golden ratio $\left(1+ \sqrt{5} \right)/2$. A quick calculation reveals that the Edwards-Anderson-type order parameters used to detect full $D_3$ symmetry breaking should then scale as $L^{-\beta}$ as well, viz. 
$m_3 = \frac{1}{L^2} \sum_{i\neq j} \bar{\mathcal{C}} \left(\left|i-j\right| \right)\approx \frac{1}{L^2} \int_0^L {\rm d}x \int_0^L {\rm d} y \left| x - y \right|^{-\beta} \propto  L^{-\beta}$, and one expects the same value of $\beta= \beta_{\rm Ising}$ only if the $D_3$-breaking transition is in the same universality class as the random Ising transition.

%\begin{align}
%m_3 &= \frac{1}{\mathcal{N}L^2} \sum_{n=1}^{\mathcal{N}}\sum_{i\neq j} \left|\bra{n}{  {\hat{\sigma}}^{\dagger}_i {\hat{\sigma}}^{\vphantom{\dagger}}_j }\ket{n}\right|^2 \\
%&\approx \frac{1}{L^2} \sum_{i\neq j} \bar{\mathcal{C}} \left(\left|i-j\right| \right) \\ 
%&\approx \frac{1}{L^2} \int_0^L {\rm d}x \int_0^L {\rm d} y \left| x - y \right|^{\varphi-2} \\ 
%&\propto L^{\varphi-2} \equiv L^{\beta}
%\end{align}
Thus, in performing finite size scaling for strong disorder, we multiply the order parameters by $L^{\beta}$ to obtain a quantity with scaling dimension zero. We display figures for the values of $\beta$ and $\nu$ that produce the best quality fit, and take the quality of this collapse as good evidence in favor of an infinite randomness critical point at $\delta_c = 0$. However, we do \emph{not} claim to extract numerically precise values of either $\nu$ or $\beta$; we only demonstrate that the data obtained are consistent with the values of these exponents one would expect from the {\it ansatz} of infinite randomness criticality. This scaling behavior is relevant only near criticality and at strong disorder (large $W$, $\delta \sim 0$). We find that the expected value of $\nu = 2$ produces a good quality fit, but so too do all $\nu \in [2,3]$. As these also satisfy the Harris bound, we cannot rule them out conclusively.

 \begin{figure}[t]
\includegraphics[width = \columnwidth]{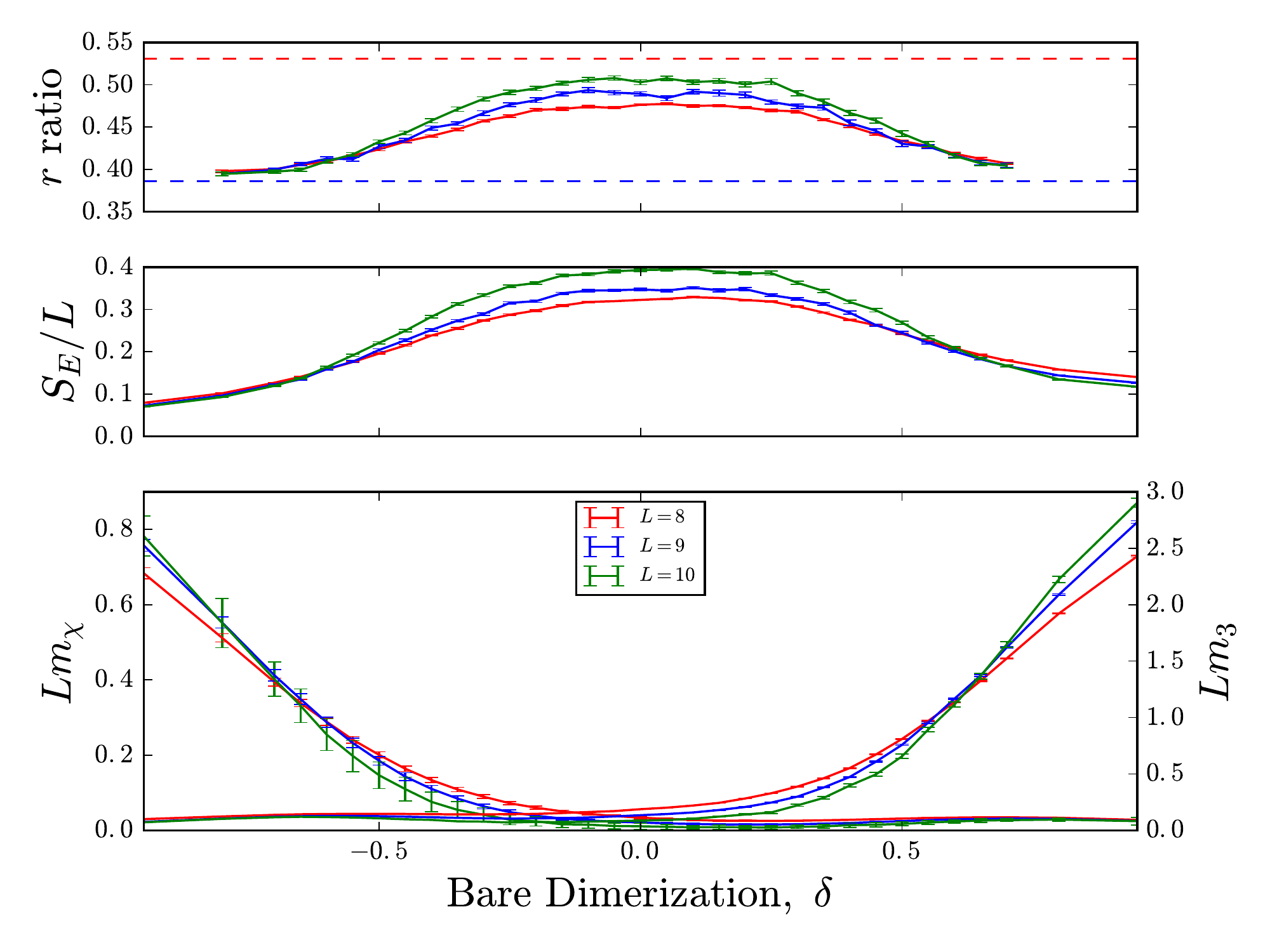} 
\caption{\label{fig:weakdisorder} Random $D_3$ chain at weak disorder, $W=0.5$.  (Top) Level statistics measured by $r$-ratio display two transitions: for $|\delta_c|\lesssim 0.5$,  $r$ tends to the ETH value $r\approx 0.53$ characteristic of the Gaussian orthogonal ensemble with increasing $L$, whereas outside this region $r\rightarrow 0.38$, indicating Poisson statistics of MBL. (Center) Half-chain entanglement entropy density  $S_{\text{E}}(L)/L$ is consistent with volume (area) law scaling in the ETH (MBL) regions. (Bottom)  Spin-glass order parameters of ${\mathbb{Z}_3}$  and chiral symmetry (scaled by $L$, see text) also show crossings at $|\delta_c|\approx 0.5$, showing that MBL coincides with the onset of symmetry-breaking.}
\end{figure}

 \begin{figure}[t]
\includegraphics[width = \columnwidth]{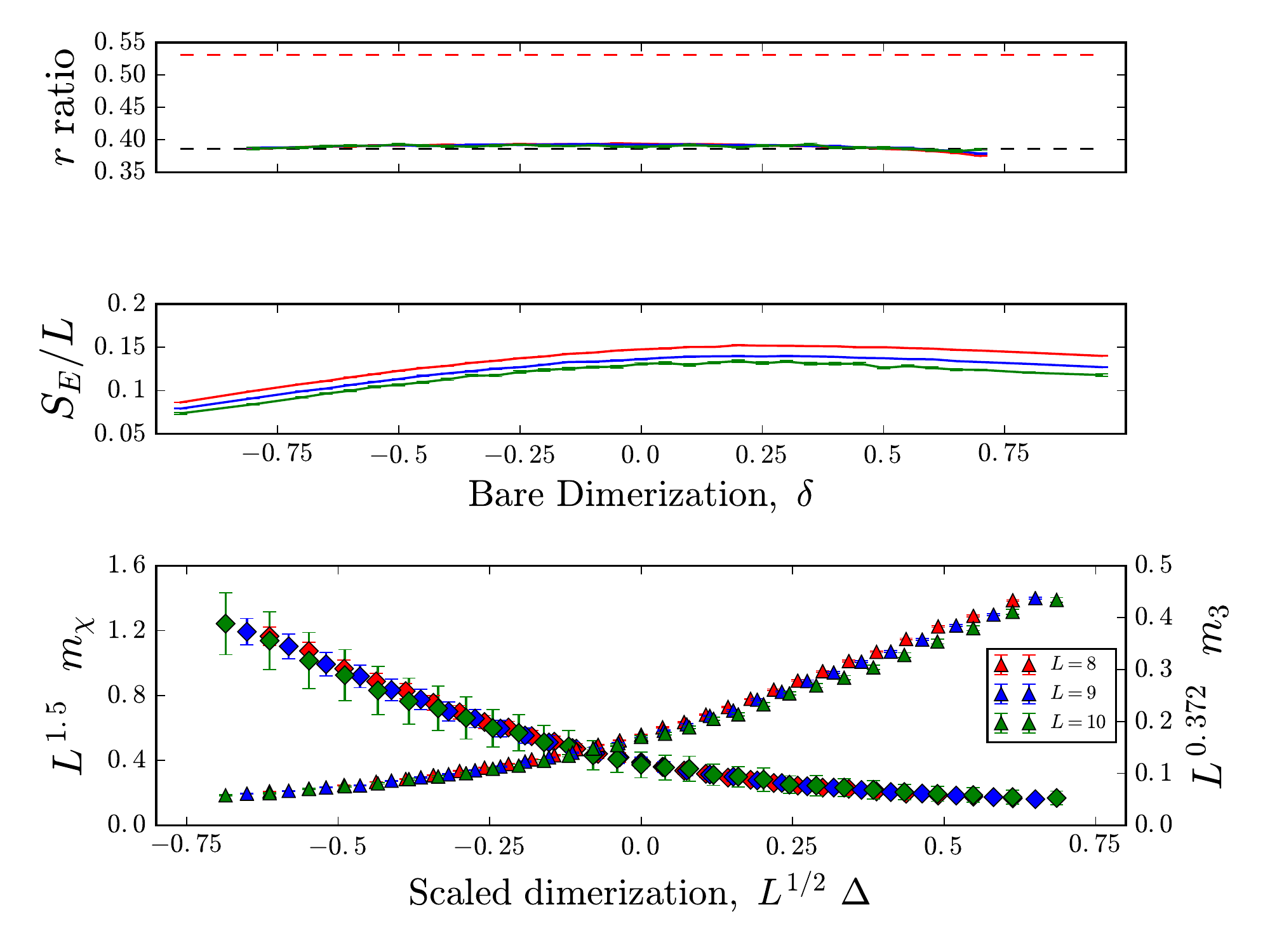}
\caption{\label{fig:strongdisorder} Random $D_3$ chain at strong disorder, $W=2.0$. (Top) Since $r\approx 0.38$ for all values of $\delta$, we infer that the system is always MBL. (Center) Entanglement entropy density is consistent with area-law scaling as $L\rightarrow\infty$, again consistent with MBL. (Bottom) Scaling collapses of  ${m_3}$ ($\triangle$) and $m_{\chi}$ ($\Diamond$), both consistent with a direct transition at $\delta_c=0$ between distinct broken-symmetry MBL phases%, controlled by an infinite-randomness critical point with $\psi=1/2$
. Here $\nu=2$, and $\Delta= \frac{\overline{\ln J} - \overline{\ln f}}{{\text{var} J}+{\text{var} f}} =\frac{1}{2W^2}\ln\frac{1+\delta}{1-\delta}$ is a rescaled tuning parameter. [The point where the two collapsed curves cross has no physical significance.]} 
\end{figure}

\section{Results}
Armed with these measures of ergodicity and symmetry breaking, we now study their behavior at weak and strong disorder.
\subsection{Weak Disorder} 
\begin{figure}[t]
\includegraphics[width = \columnwidth]{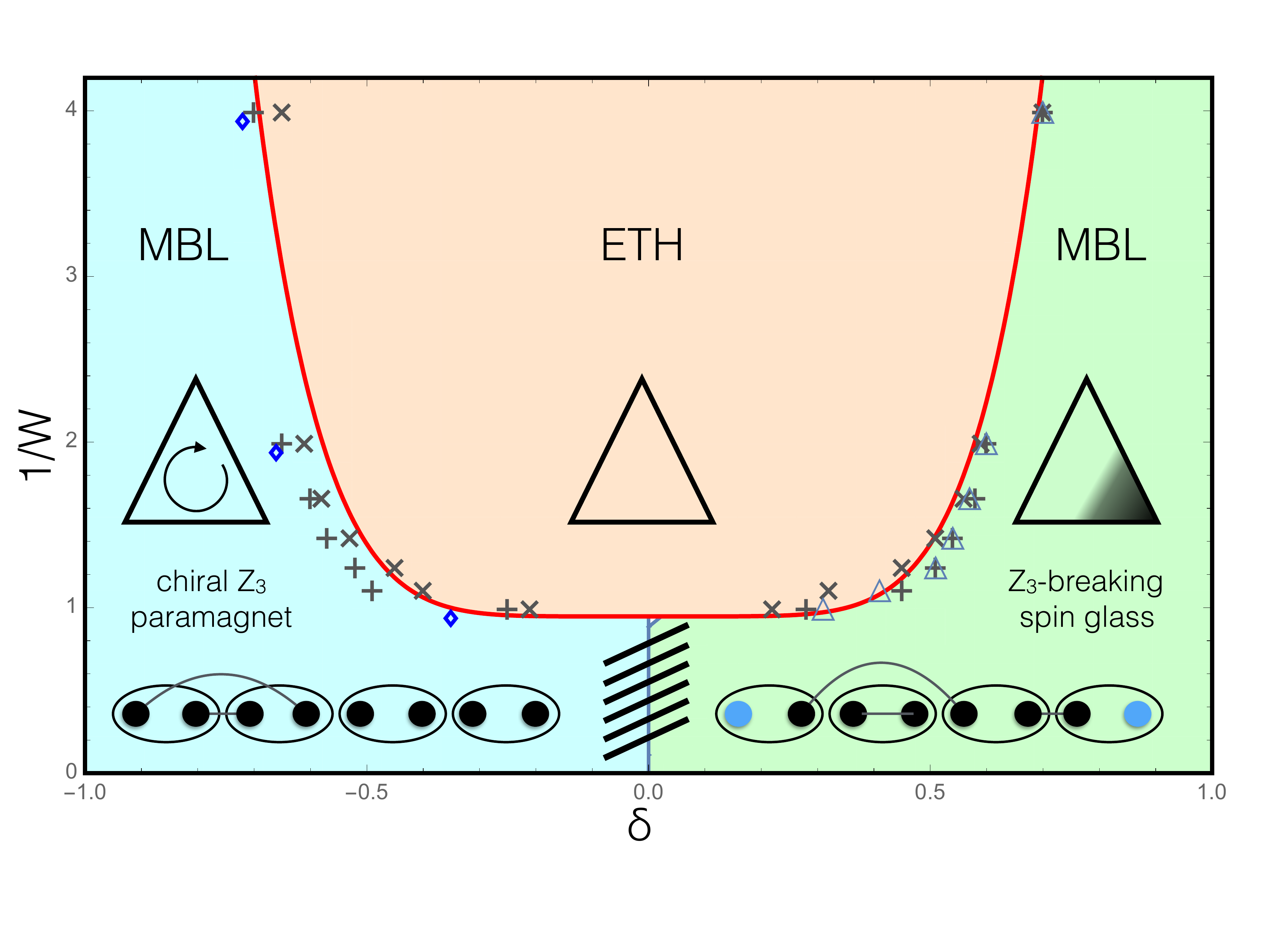} 
\caption{\label{fig:phasediag} Non-equilibrium global phase diagram of random $D_3$ chain. The MBL-ETH boundary (red line) is an estimate based on crossings in level statistics and scaled entanglement entropy (denoted $\times$, $\dagger$ respectively). For weak disorder we also indicate crossings in the ${\mathbb{Z}_3}$ ($\triangle$) and chiral ($\Diamond$) order parameters. At strong disorder, we find scaling collapse consistent with an infinite-randomness critical point at $\delta_c=0$; however, we cannot conclusively rule out a non-ergodic quantum critical glass in the transition region (hatched). 
{(Inset) schematic of the symmetry-breaking pattern and the topological/trivial phases of dual parafermions.}
}
\end{figure}

For a representative choice of weak disorder, $W=0.5$ (Fig.~\ref{fig:weakdisorder}), we identify a pair of transitions both in level statistics and entanglement. For $|\delta| \lesssim 0.5$, the $r$-ratio increases towards the ETH value of 0.53 with increasing system size, whereas outside this region it decreases towards the MBL value of 0.38. This is also consistent with the change from area-law to volume-law scaling observed in the eigenstate-averaged entanglement entropy ($S_E = \sum_{n=1}^{\mathcal{N}}S_E^{(n)}/\mathcal{N}$). We conclude that an ETH region for $|\delta| \lesssim 0.5$ is flanked by a pair of MBL phases. We next study symmetry-breaking, by considering the scaled quantities $L{m_3}, Lm_{\chi}$: these scale $\sim L$ in a phase with spin-glass order, and vanish in a symmetry preserving phase. Crossings of curves of either $L{m_3}$ or $Lm_{\chi}$ corresponding to different system sizes approximately locate transitions between a paramagnetic and broken-symmetry phase. Within the accuracy of our numerics, these appear to coincide with the crossings in level statistics, with the $\delta\gtrsim 0.5$ ($\delta\lesssim 0.5$) phase breaking the $D_3$ (${\mathbb{Z}_2}$) symmetry in a spin glass sense. We track similar behavior up to $W\approx1.0$, whereupon the ETH phase disappears; we therefore identify $W>1.0$ with strong disorder. Fig.~\ref{fig:phasediag} shows the extent of the ETH phase and approximate locations of the crossings in $r, {m_3}$ and $m_\chi$. Crucially, a fully $D_3$ symmetric MBL phase is absent, in accord with general symmetry restrictions~\cite{PhysRevB.94.224206}.

\subsection{Strong Disorder} For strong disorder (e.g. $W=2.0$, see Fig.~\ref{fig:strongdisorder}) we find no evidence for an ETH phase in either level statistics or entanglement: with increasing $L$, $r\rightarrow 0.385$ and $S_{E}/L$ decreases, consistent with either MBL or eigenstate criticality~\cite{QCGPRL}, for all $\delta$. Extrapolating from weak disorder, we see that the two MBL-ETH transition lines appear to converge at around $W\approx 1.0$. Recall that the two  MBL phases have distinct broken symmetries. At strong disorder, as the ETH phase is absent and an MBL paramagnet is inconsistent with $D_3$ symmetry~\cite{PhysRevB.94.224206}, there must be either a direct transition between the two broken-symmetry MBL phases, or an intervening symmetry-preserving quantum critical glass phase. [A third possibility, namely that a narrow sliver of ETH phase persists to strong disorder~\cite{deRoeckHuveneers,deRoeckImbrie, HusePersonal}, is not evident in our numerics but we cannot rule it out for $L\rightarrow \infty$.] Although it is challenging to distinguish these scenarios given our limited range of system size{s}, we find some support for the former via finite-size scaling analysis, as follows. First, we argue that self-duality of \eqref{eq:Ham} implies symmetry under reflection of $\delta$, fixing a single direct transition to occur at $\delta_c=0$ (we ignore small finite-size corrections to duality from the open boundary conditions; note that the weak disorder {MBL-ETH} transitions are at {roughly} $\pm \delta_c$, consistent with duality). Second, we assume that a direct transition between MBL phases is controlled by an infinite-randomness fixed point, with length-time scaling $\ell^\psi \sim \ln t$. We fix $\psi=1/2$, characteristic of a random singlet critical point, the generic scenario~\cite{SenthilPotts} in systems with Abelian global symmetries (as both phases lack  full non-Abelian $D_3$ symmetry), and use the scaling form
\be\label{eq:scalingform}
 m(\delta, L) = L^{-\beta} \Phi\left[{L}^{1/\nu}\Delta\right],
\ee
where $\nu = 2$, $\Delta = \frac{\overline{\ln J} - \overline{\ln f}}{{{\text{var} J}+{\text{var} f}}} =\frac{1}{2W^2}\ln\frac{1+\delta}{1-\delta}$ tunes across the $\delta_c=0$ critical point, and $\Phi$ is a universal function. For $m_3$,  the exponent $\beta = 0.372$ derives from known results for $\psi=1/2$ infinite-randomness critical scaling of the random transverse-field Ising model~\cite{FisherRSRG1,FisherRSRG2}, which predict an exponent $\beta_{\rm Ising} = 2 -\varphi$. The same scaling does not hold for $m_{\chi}$, where $\beta$ is not known analytically, and we merely fit these data using the scaling form \eqref{eq:scalingform}. As we see from the bottom panel of Fig.~\ref{fig:strongdisorder}, the data for ${m_3}, m_\chi$ show reasonably good collapse when scaled according to \eqref{eq:scalingform}, with the exponent $\beta_{\chi} = 1.5$ chosen to show that a satisfactory collapse according to \eqref{eq:scalingform} exists, though we do \emph{not} claim to have extracted a precise value from the data. With no other free parameters, the collapses are at least consistent with a direct infinite-randomness transition between the ${\mathbb{Z}_3}$ spin glass and the chiral paramagnet (though other exponent choices also give  reasonable data collapses). Furthermore, we cannot rule out a possible sliver of a quantum critical glass {\it phase} rather than a critical point, which might show similar scaling collapse for accessible system sizes. We indicate this ambiguity by the hatched lines marking the transition region in Fig.~\ref{fig:phasediag}.

\section{Discussion} Our results are summarized in a global non-equilibrium phase diagram for the random $D_3$ chain (Fig.~\ref{fig:phasediag}). To the extent we are able to determine, MBL always coincides with the breaking of the non-Abelian discrete $D_3$ symmetry. The MBL ${\mathbb{Z}_3}$ spin glass completely breaks this symmetry; in a different language, this phase will host parafermionic zero modes stable at finite temperature. The other MBL phase is a ${\mathbb{Z}_3}$-symmetric paramagnet, which --- unlike its ground-state counterpart --- breaks the remaining ${\mathbb{Z}_2}$ chiral symmetry,  consistent with the no-go theorem forbidding MBL with non-Abelian symmetry. 
%%% AARON: commentary inserted in response to feedback from Ref A:
At $\delta = -1$, the model describes a trivial paramagnet, the eigenstates of which feature extensive degeneracies due to the $D_3$ symmetry. Our numerics show that for any finite interaction $\delta > -1$, the degeneracy due to the chiral $\mathbb{Z}_2$ component of $D_3$ is lifted by spontaneous symmetry breaking. This is precisely the instability of the MBL phase to non-Abelian symmetries predicted in a previous work\cite{PhysRevB.94.224206}. While an ETH phase intervenes at weak disorder, at strong disorder we find evidence for an infinite-randomness transition between these distinct broken-symmetry MBL phases, although a  more exotic possibility, an athermal quantum critical glass, cannot be definitively ruled out. We conjecture that similar features apply, {\it mutatis mutandis}, to other non-Abelian random spin chains with $D_n$ symmetry. Further investigation of such models, e.g. via matrix product state methods~\cite{PhysRevB.94.041116,PhysRevLett.116.247204, PhysRevLett.118.017201,PhysRevB.95.035116}, would be an interesting avenue for future work.

\begin{acknowledgements} We thank  A.~Prakash, {W.-W.~Ho}, V.~Khemani and  D.~Abanin for discussions. This research was supported by the National Science Foundation via Grants DGE-1321846 (Graduate Research Fellowship Program, A.J.F.),  DMR-1455366 (S.A.P.) and PHY-1125915 (at the KITP, A.J.F, R.V., S.A.P.), the Department of Energy via the LDRD program of LBNL (R.V.), and by NSF DMR-1653007 (A.C.P.). 
\end{acknowledgements}

\begin{appendix}

\section{Order Parameters} \label{sec:OPs}
\subsection{Subsidiary $\mathbb{Z}_2$ breaking in the $\mathbb{Z}_3$ spin glass}
%\subsection{Details of the $\mathbb{Z}_3$ order parameter} 
\label{subsec:subsidZ2}
For the parameter space we investigate, the model \eqref{eq:Ham} has a global $D_3 = S_3$ symmetry generated by the action of a $\mathbb{Z}_3$ cycle ${\hat{\mathcal{Q}}} = \prod_j {\hat{\tau}}_j$ and a $\mathbb{Z}_2$ swap ${\hat{\mathcal{X}}} \equiv \prod_j {\hat{\mathcal{X}}}_j$, where ${\hat{\mathcal{X}}}_j$ interchanges local eigenstates $\ket{1} \leftrightarrow \ket{2}$ of \emph{either} $\sigma_j$ or $\tau_j$, while leaving $\ket{0}$ invariant. In the $\mathbb{Z}_2$ Ising model, spin glass order is detected using an Edwards-Anderson order parameter. The Ising chain has only a global $\mathbb{Z}_2$ symmetry generated by $ {\hat{\mathcal{S}}} = \prod_j \hat{\sigma}^x_j$, and the corresponding eigenstate-averaged order parameter is
\be  m_{\rm Ising} = \frac{1}{\mathcal{N}L^2} \sum_{n=1}^{\mathcal{N}}\sum_{i\neq j} \left|\bra{n}{\hat{\sigma}^z_i \hat{\sigma}^z_j }\ket{n}\right|^2, \ee
where $\hat{\sigma}^z_j$ anticommutes with the local symmetry generator, ${\hat{\sigma}}^x_j$. For $D_3$, the analogous operator ${\hat{\mathcal{J}}}_j = \frac{1}{i \sqrt{3}} \left( {\hat{\tau}}^{\vphantom{\dagger}}_j - {\hat{\tau}}^{\dagger}_j \right)$ anticommutes with the chiral $\mathbb{Z}_2$ generator ${\hat{\mathcal{X}}}_j$, as evinced by 
their matrix representations in the $\hat{\tau}^{\vphantom{\dagger}}_j$-basis:
\be  {\hat{\mathcal{X}}}_j = \begin{pmatrix} 1 & 0 & 0 \\ 0 & 0 & 1 \\ 0 & 1 & 0 \end{pmatrix}_j~,\quad  {\hat{\mathcal{J}}}_j = \begin{pmatrix} 0 & 0 & 0 \\ 0 & 1 & 0 \\ 0 & 0 & {\rm-}1 \end{pmatrix}_j ,
\ee
which have the appropriate Pauli matrix structure. We therefore use the conjugate operator ${\hat{\mathcal{J}}}$ to construct the chiral Edwards-Anderson order parameter for ${\hat{\mathcal{X}}}$-breaking, 
\be m_{\chi} = \frac{1}{\mathcal{N}L^2} \sum_{n=1}^{\mathcal{N}}\sum_{i\neq j} \left|\bra{n}{\hat{\mathcal{J}}}_i {\hat{\mathcal{J}}}_j \ket{n}\right|^2. \ee

As previously stated, ${\hat{\mathcal{X}}}_j$ has the same matrix form in the $\sigma$-basis; one may also construct the conjugate in that basis, ${\hat{\mathcal{K}}}_j = \frac{1}{i \sqrt{3} } \left( {\hat{\sigma}}^{\vphantom{\dagger}}_j - {\hat{\sigma}}^{\dagger}_j \right)$, whereupon the order parameter becomes
\begin{align}
\tilde{m}_{\chi} &= \frac{1}{\mathcal{N}L^2} \sum_{n=1}^{\mathcal{N}} \sum_{i\neq j} \left| \bra{n} {\hat{\mathcal{K}}}_i {\hat{\mathcal{K}}}_j \ket{n}\right|^2 \\
&= \frac{1}{9 \mathcal{N}L^2} \sum_{\substack{n \\ i \neq j}} \left|\bra{n} {  \left( {\hat{\sigma}}^{\dagger}_i {\hat{\sigma}}^{\vphantom{\dagger}}_j  +  {\hat{\sigma}}^{\vphantom{\dagger}}_i {\hat{\sigma}}^{\dagger}_j  - {\hat{\sigma}}^{\dagger}_i {\hat{\sigma}}^{\dagger}_j -  {\hat{\sigma}}^{\vphantom{\dagger}}_i  {\hat{\sigma}}^{\vphantom{\dagger}}_j   \right)}\ket{n}\right|^2  .\label{eq:sigz2full}
\end{align}
However, the energy eigenstates $\ket{n}$ are constructed as eigenstates of ${\hat{\mathcal{Q}}} = \prod_j {\hat{\tau}}_j$, and therefore correspond to states with a fixed $\mathbb{Z}_3$ charge $Q= \sum_j q_j ~\tmod{3}$. The latter two terms in \eqref{eq:sigz2full} both change the total $\mathbb{Z}_3$ charge by $\pm 1$, and therefore have a trivially zero expectation value. The $\sigma$-basis order parameter then becomes $\tilde{m}_{\chi} = \frac{1}{9 \mathcal{N}L^2} \sum_{n=1}^{\mathcal{N}}\sum_{i\neq j} \left|\bra{n}{  \left( {\hat{\sigma}}^{\dagger}_i {\hat{\sigma}}^{\vphantom{\dagger}}_j  +  {\hat{\sigma}}^{\vphantom{\dagger}}_i {\hat{\sigma}}^{\dagger}_j\right)}\ket{n}\right|^2 =  \frac{4}{9} m_3$,
%\begin{align}
%\tilde{m}_{\chi} &= \frac{1}{9 \mathcal{N}L^2} \sum_{n=1}^{\mathcal{N}}\sum_{i\neq j} \left|\bra{n}{  \left( {\hat{\sigma}}^{\dagger}_i {\hat{\sigma}}^{\vphantom{\dagger}}_j  +  {\hat{\sigma}}^{\vphantom{\dagger}}_i {\hat{\sigma}}^{\dagger}_j\right)}\ket{n}\right|^2 %\\
%%&= \frac{4}{9 \mathcal{N}L^2} \sum_{n=1}^{\mathcal{N}}\sum_{i\neq j} \left|\bra{n}{  {\hat{\sigma}}^{\dagger}_i {\hat{\sigma}}^{\vphantom{\dagger}}_j }\ket{n}\right|^2  \\
%%&
%=  \frac{4}{9} m_3, \end{align}
i.e, proportional to the Edwards Anderson order-parameter for $\mathbb{Z}_3$-breaking. As an aside, using either of the other two $\mathbb{Z}_2$ symmetry operators ${\hat{\mathcal{X}}\hat{\mathcal{Q}}}$, ${\hat{\mathcal{X}}\hat{\mathcal{Q}}}^2$ and constructing the corresponding order parameters in the $\sigma$-basis has the same result: the only respective changes are factors of $\omega$ and $\omega^2$ multiplying the trivial ${\hat{\sigma}}^{\dagger}_i {\hat{\sigma}}^{\dagger}_j  $ term. Hence, the $\mathbb{Z}_3$-breaking spin glass necessarily breaks chiral $\mathbb{Z}_2$, and so $D_3$ is fully broken in this phase.

% \subsection{Auxiliary $\mathbb{Z}_2$ order parameter} \label{subsec:auxZ2} 
\subsection{Details of the auxiliary $\mathbb{Z}_2$ order parameter} \label{subsec:auxZ2} 
The $\mathbb{Z}_2$ order parameter defined in \eqref{eq:z2op} measures chiral order in the $\tau$-basis -- i.e., breaking of $D_3$ to a $\mathbb{Z}_3$-preserving paramagnet. 
%%%%AARON: The following was added following feedback from Referee A
 This order parameter is designed for use only in a paramagnetic phase, wherein $\mathbb{Z}_3$ is preserved, and is not a useful measure of $\mathbb{Z}_2$ breaking when evaluated with states that are \emph{not} eigenstates of the $\mathbb{Z}_3$ generator. Although the derivation from conjugacy to the $\mathbb{Z}_2$ generator derived in the preceding subsection is sufficient, we also performed several sanity checks to confirm this quantity is reasonable. To wit, it has the appropriate action on generic states constructed specifically to preserve or break the $\mathbb{Z}_2$ symmetry. Additionally, it is everywhere zero when computed in \emph{ground states} of the $D_3$ model, where the quantum phase transition at $T=0$ only admits either a fully $D_3$-preserving paramagnet or a spin glass that breaks $D_3$ completely. Lastly, it shows chiral ordering when calculated in eigenstates of the chiral $\mathbb{Z}_3$ Hamiltonian, which breaks explicitly the $\mathbb{Z}_2$ subgroup of $D_3$ (provided $\bar{f_j} > \bar{J_j}$). The fact that this quantity is zero even in $D_3$ ground states and all $\mathbb{Z}_3$ eigenstates for $\bar{J_j}>\bar{f_j}$ confirms that it is only meaningful on the putatively paramagnetic side of the phase diagram, $\delta \leq 0$.

\section{Real space renormalization group}
The real-space renormalization group (RSRG) is a perturbative decimation scheme originally formulated to construct and study approximate ground states of disordered spin chains. The RG generates a flow in the space of couplings, which is asymptotically exact if towards stronger disorder. Properties of the ground state are then controlled by an `infinite-randomness' fixed point with universal scaling properties. RSRG-X is an extension of this approach that targets excited states\cite{PekkerRSRGX,PhysRevB.95.024205}. In each step of RSRG-X, we diagonalize the strongest remaining term in the Hamiltonian, $H_\Omega$, ignoring all other terms. In systems with an Abelian global symmetry (e.g $\mathbb{Z}_n$), eigenstates of  $H_\Omega$ will generically be non-degenerate singlets. Choosing any of these states then completely fixes the states of the `decimated' spins in $H_\Omega$, whose virtual fluctuations mediate interactions between the remaining spins, which may be computed via second-order perturbation theory. A specific temperature is targeted by weighting these choices with appropriate Boltzmann probabilities.

\subsection{Real space renormalization group for $D_3$} 
\label{subsec:s3rg}
This procedure is complicated by the presence of non-Abelian symmetries, since diagonalizing the strong bond may not completely determine the states of the spins involved. Instead, they may collectively transform according to a $\mathcal{D}>1$- dimensional irreducible representation of the symmetry group. In the RG language, this means a `superspin' survives the decimation, and couples to the remaining spins computed via first-order perturbation theory%, i.e. mediated by direct, rather than virtual, processes
. While such non-singlet decimations can sometimes be ruled out in ground states -- as with the random-bond Heisenberg antiferromagnet  -- they will always occur in finite-energy density excited states. Such a finite density of {\it localized} non-singlet states in turn implies an exponential degeneracy. Absent fine-tuning, this will either drive the system thermal (in the ETH sense), or else be split by spontaneously breaking the non-Abelian symmetry down to an Abelian subgroup\cite{XXZPaper,PhysRevB.94.224206}
. 

Athermal phases that lack a local tensor product structure   --- dubbed `quantum critical glasses' (QCGs)  --- evade this argument as their symmetry-enforced degeneracies cannot be  lifted by local perturbations; therefore QCGs may be paramagnetic, though their eigenstate entanglement entropy will exceed the area law characteristic of MBL. The key question is to determine which of these distinct possibilities --- an ETH phase, a QCG, or a broken-symmetry MBL phase --- emerges for a given choice of parameters. One  approach is to keep higher-order contributions in perturbation theory when decimating spins; these could indicate, e.g., a bias towards thermalization or symmetry-breaking. This was successfully employed in the XXZ chain\cite{XXZPaper}
, where the leading corrections drive spin glass order. In the $S_3$ case, however, this approach is challenging since the reduced symmetry permits many distinct competing higher-order terms, with no clear dominant contribution.

A related issue arises even in the  nominally simpler $\mathbb{Z}_3$ case, where $\phi_i$ also takes random values in $[0,2\pi)$. Na\"{i}vely, every decimation in this case should result in a singlet; however, for values of $\phi_j \approx 0 ~\text{mod}{\pi/3}$, rather than an $O(1)$ energy splitting between the different singlets set by the overall scale of $H_\Omega$, two of the singlets are nearly degenerate. Even if avoided at the initial stages of the RG, such near-degeneracies become more frequent as the RG proceeds, since the couplings are multiplicatively renormalized. Therefore, even in the $\mathbb{Z}_3$ case, RSRG-X becomes challenging, necessitating a different approach. For completeness, we briefly discuss RSRG-X in the $\mathbb{Z}_3$ case, to see how such degeneracies appear.

\subsection{Real Space Renormalization Group for $\mathbb{Z}_3$}

We begin with the Hamiltonian from the main text,
\be H = \sum_{j=1}^{L-1} J_j e^{i \phi_j} \sigma^\dagger_j \sigma^{\vphantom{\dagger}}_{j+1} + \sum_{j=1}^{L} f_j e^{i \theta_j} \tau^{\vphantom{\dagger}}_j +h.c., \ee
and rewrite in terms of parafermion operators $\pfC_j$, defined via % are related to the above by:
\be \sigma^{\dagger}_j \sigma^{\vphantom{\dagger}}_{j+1} = \omega^{-1} \pfCconj_{2j} \pfC_{2j+1}, \quad \tau_j = \omega^{-1} \pfCconj_{2j-1} \pfC_{2j}.  \ee
It is convenient to redefine the parameters via
\be \left( K_x~,~\varphi_x \right) = \begin{cases} \left( J_j~,~\phi_j \right) & \text{if } x = 2j \notag \\ \left( f_j~,~\theta_j \right) & \text{if } x = 2j-1 \notag \end{cases} \ee
The Hamiltonian then takes the simple form:
\be H = \frac{1}{\omega} \sum_{x=1}^{2L-1}  K_x \left( e^{i \varphi_x} \pfCconj_x \pfC_{x+1} + e^{-i \varphi_x} \pfC_x \pfCconj_{x+1} \right). \ee
Note that this procedure is analogous to rewriting Ising spins in terms of Majorana fermion operators: each clock spin has been replaced by a pair of parafermions. This is convenient since both `transverse field' and `bond' terms in the clock model are `bond` terms in the parafermion language, permitting us to discuss them on the same footing. We now proceed to implement the RSRG procedure.
\begin{figure}[t]
\includegraphics[width=0.8\columnwidth]{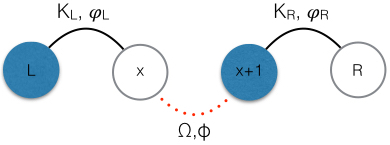}
\caption{\label{fig:z3rg} {\it Bond decimation for $\mathbb{Z}_3$.} The two parafermion modes corresponding to the bond drop out of the chain, and the resulting couplings leave the Hamiltonian self-similar, with one fewer site to consider in subsequent steps of the RG.}
\end{figure}

In this language, the strongest remaining term in a given RG step, $H_{\Omega}$, corresponds to some [pseudo-]site $x$, and writing $K_x$, $\phi_x \rightarrow \Omega, \Phi$ on the strong bond,
\be H_{\Omega} = H_{x,x+1}=\frac{\Omega}{\omega} \left( e^{i \Phi} \pfCconj_x \pfC_{x+1} + e^{- i \Phi} \pfC_x \pfCconj_{x+1} \right). \ee
The operator $\omega^{-1} \pfCconj_x \pfC_{x+1} $ recovers either $ \tau_j $ or $\sigma^{\dagger}_j \sigma^{\vphantom{\dagger}}_{j+1}$, and therefore has eigenvalues $\omega^{q_x}$, where $q_x$ is a book-keeping integer with value $0, 1, 2$. Thus, the unperturbed energy is $ \mathcal{E}_0  = 2 \Omega {\rm cos} \left( \Phi + \frac{2 \pi q_x}{3} \right)$.
The perturbing potential $V$ consists of couplings to nearest neighbors, so we need only consider four pseudo-sites: $x,x+1$, and the two neighboring sites on the left/right, denoted $L/R$ respectively (Fig.~\ref{fig:z3rg})

The dashed red line in Fig.~\ref{fig:z3rg} indicates the strongest bond, with coupling $\Omega$, and the two solid lines are contained in the perturbing potential, $V$: 
\begin{align} V \equiv \frac{\Omega}{\omega} \left(\{ K_L \left(e^{i \varphi_L} \pfCconj_L \pfC_x +  e^{- i \varphi_L}  \pfC_L \pfCconj_x   \right) + \right. \notag \\
\left. K_R \left( e^{i \varphi_R} \pfCconj_{x+1} \pfC_{R}+ e^{- i \varphi_R}  \pfC_{x+1} \pfCconj_R  \right) \right\}, \end{align}

so that the full Hamiltonian is given by \be H = H_0 + \lambda V~, \quad \lambda = \Omega^{-1}. \ee
Next we implement degenerate perturbation theory, and we have 
\be \matel{\psi_{\alpha}}{\tilde{V_1}}{\psi_{\beta}} = \Omega^{-2} \sum_{\ket{\gamma} \notin \mathcal{H}_d}  \frac{ \matel{\varphi_{\alpha}}{V}{\gamma} \matel{\gamma}{V}{\varphi_{\beta}} }{\mathcal{E}_d - \matel{\gamma}{H_0}{\gamma}}. \ee

The result is that the parafermions on sites $x$ and $x+1$ drop out of the chain, frozen in the state $\mathcal{E}_0$, and a new term is coupling between the parafermions $L$ and $R$ with magnitude $\tilde{K}_{LR}$ and chiral phase $\tilde{\varphi}_{LR}$ is generated via
\begin{align} \tilde{K}_{LR} &= { \frac{K_L K_R}{\Omega } \frac{ {\rm cos} \left( \Phi + \frac{2 \pi q_x}{3} \right) }{ {\rm cos~} 2\left(\Phi + \frac{2 \pi q_x}{3} \right) + \frac{1}{2} }  } \\
\tilde{\varphi}_{LR} &= { \varphi_L + \varphi_R- \frac{2 \pi q_x}{3} }.
\end{align}
Note that the effective coupling diverges if the random phase $\Phi$ is an integer multiple of $\pi$, corresponding to a non-chiral bond, for which there are two degenerate states (the generic bond in the $S_3$ case). For an infinite $\mathbb{Z}_3$ chain, at some point in the RG the effective coupling resulting from a decimation will be close enough to an integer multiple of $\pi$ to produce a resonance, and effectively an $S_3$ Potts term. Hence RSRG, while nominally controlled for the $\mathbb{Z}_3$ case, may have lead to dangerous resonances during the RG flow.

\section{Additional numerical evidence}

\subsection{Weak Disorder}

\begin{figure}[t!]
\caption{$W=0.5$ (Fig. 1 of main text)}{
\includegraphics[width=0.95\columnwidth]{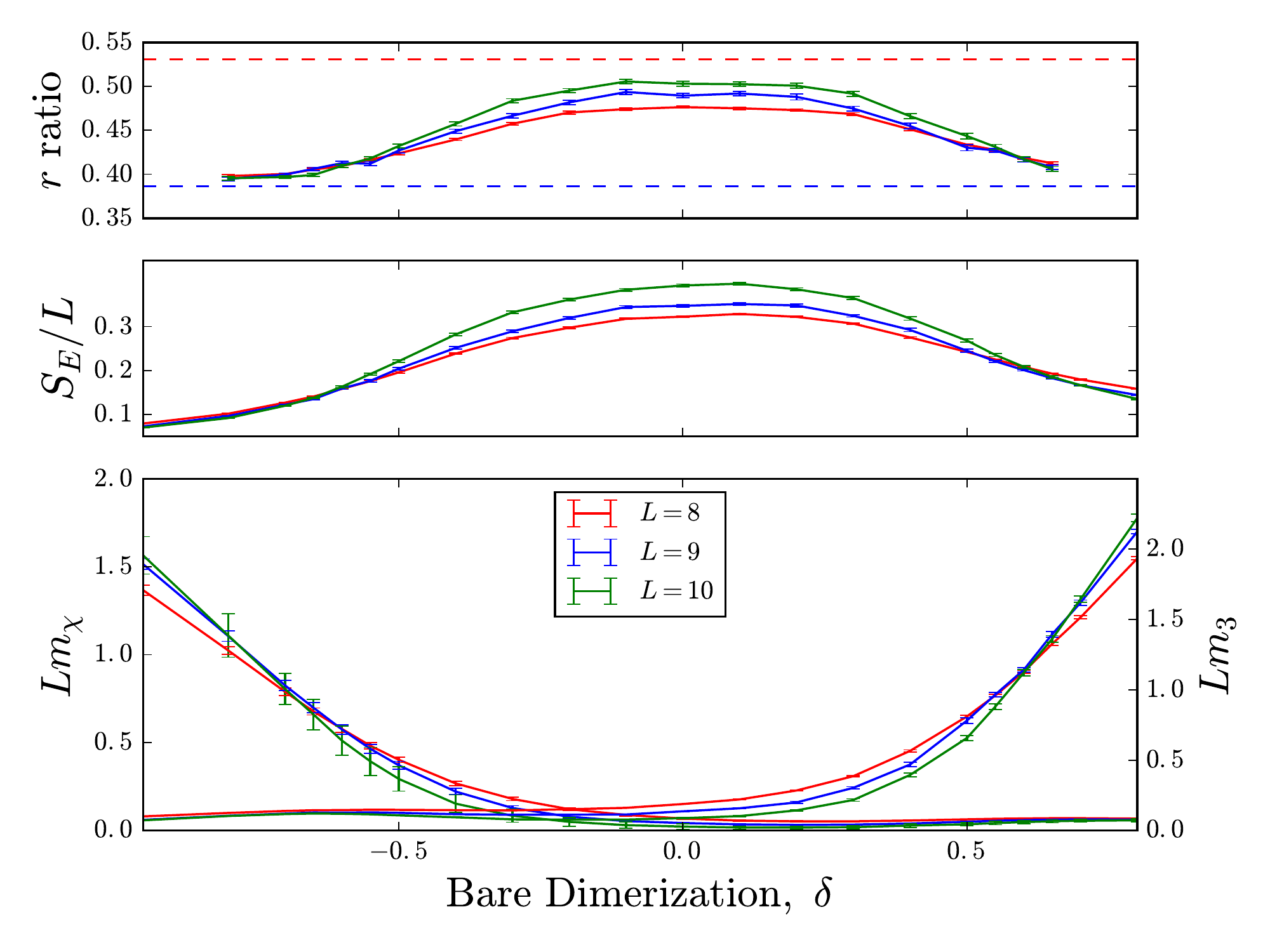}
\label{fig:subfig2wk}}
\end{figure}

\begin{figure}[t!]
\caption{Additional results for $W=0.6$}{
\includegraphics[width=0.95\columnwidth]{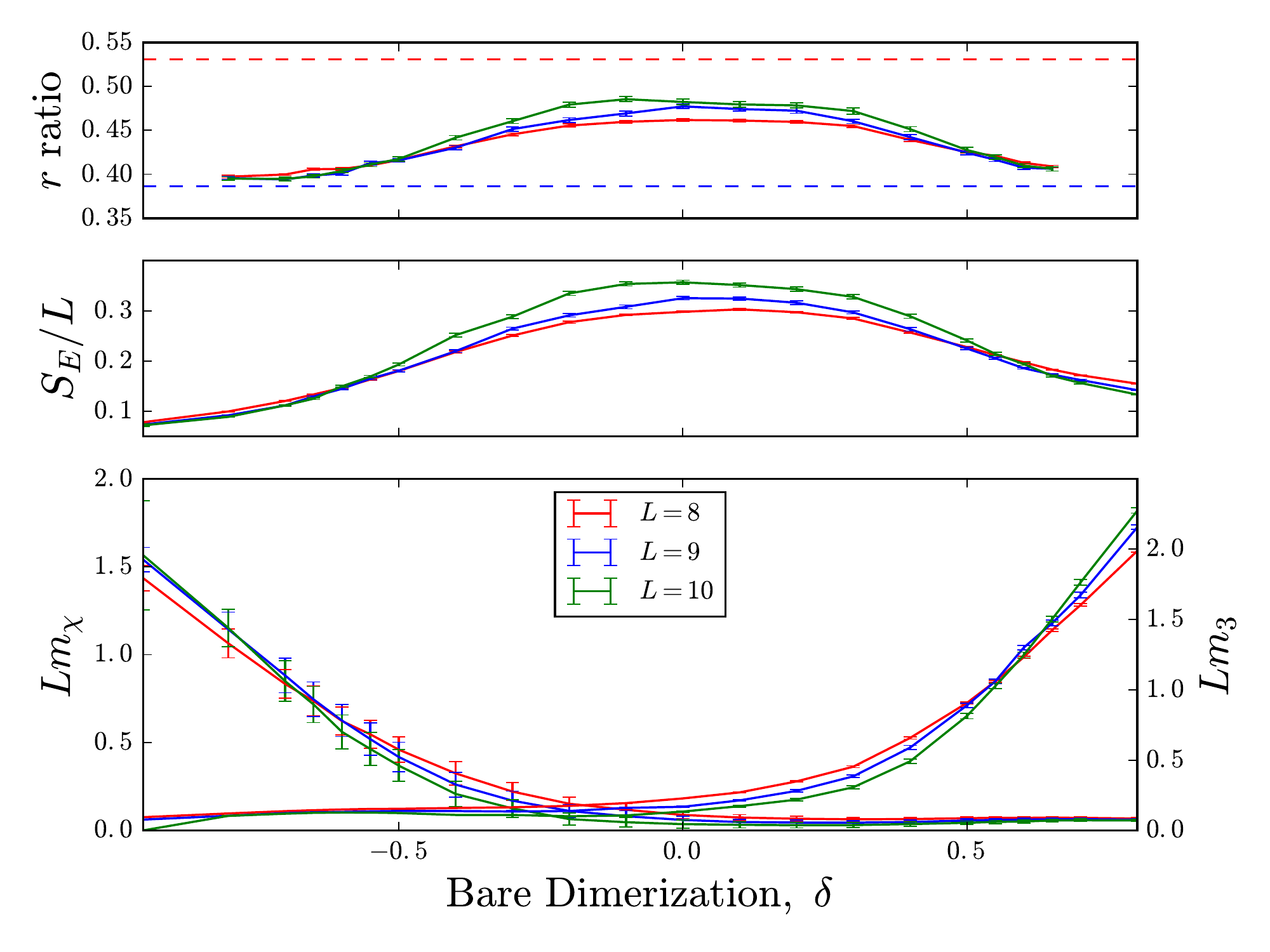}
\label{fig:subfig3wk}}
\end{figure}

\begin{figure}[t!]
\caption{Additional results for $W=0.7$}{
\includegraphics[width=0.95\columnwidth]{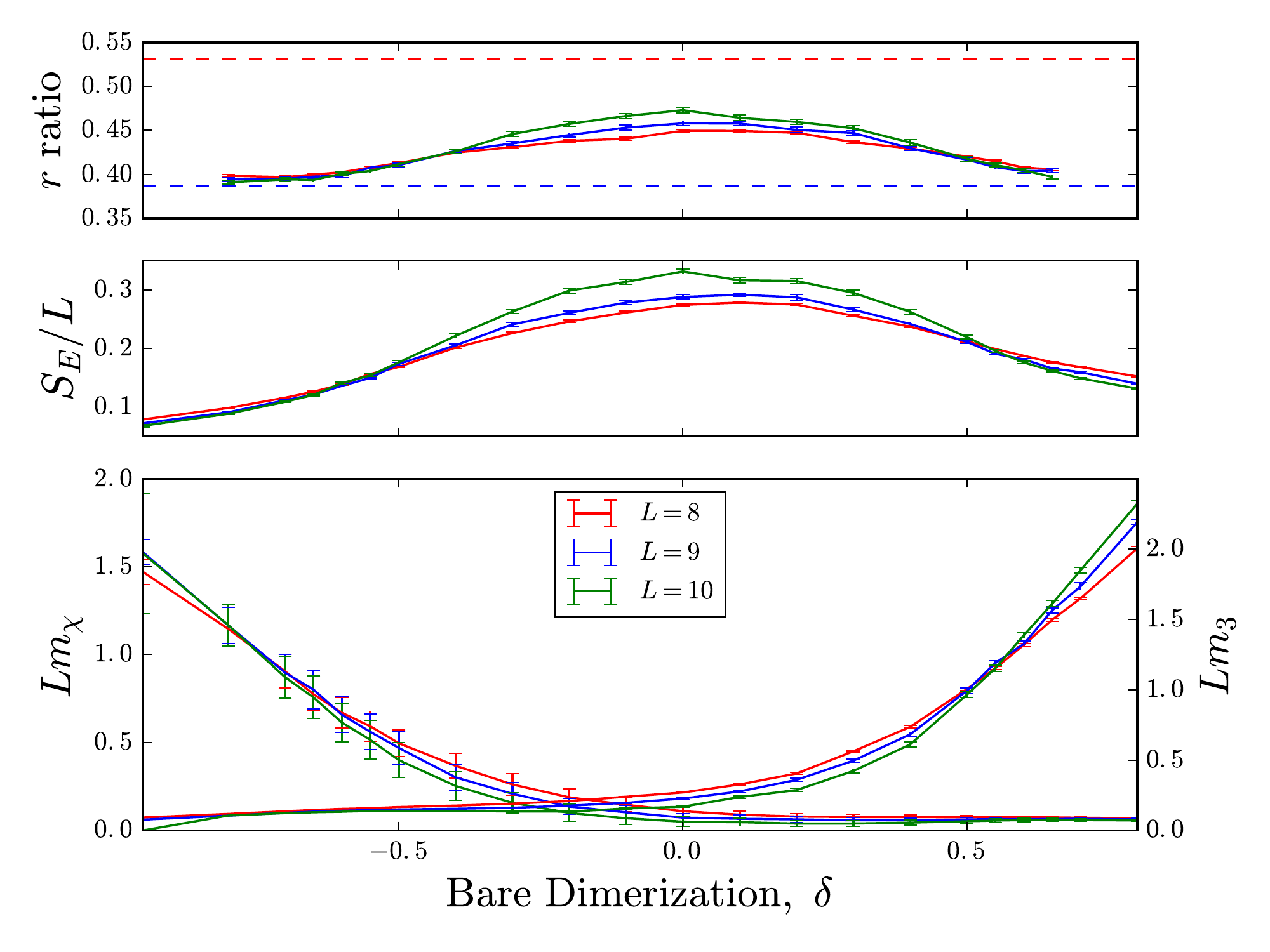}
\label{fig:subfig4wk}}
\end{figure}

\begin{figure}[t!]
\caption{Additional results for $W=0.8$}{
\includegraphics[width=0.95\columnwidth]{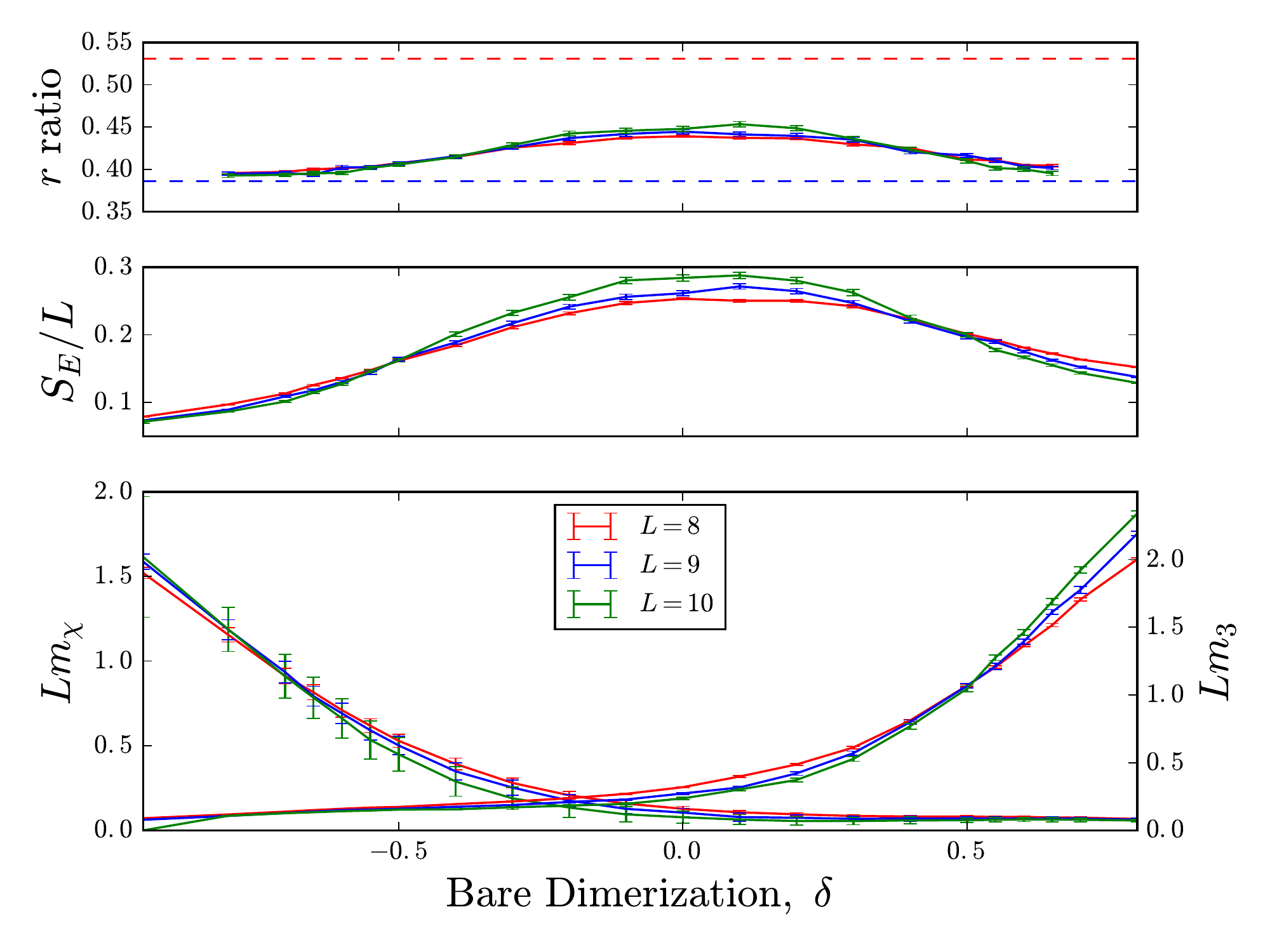}
\label{fig:subfig5wk}}
\end{figure}

\begin{figure}[t!]
\caption{Additional results for $W=0.9$}{
\includegraphics[width=0.95\columnwidth]{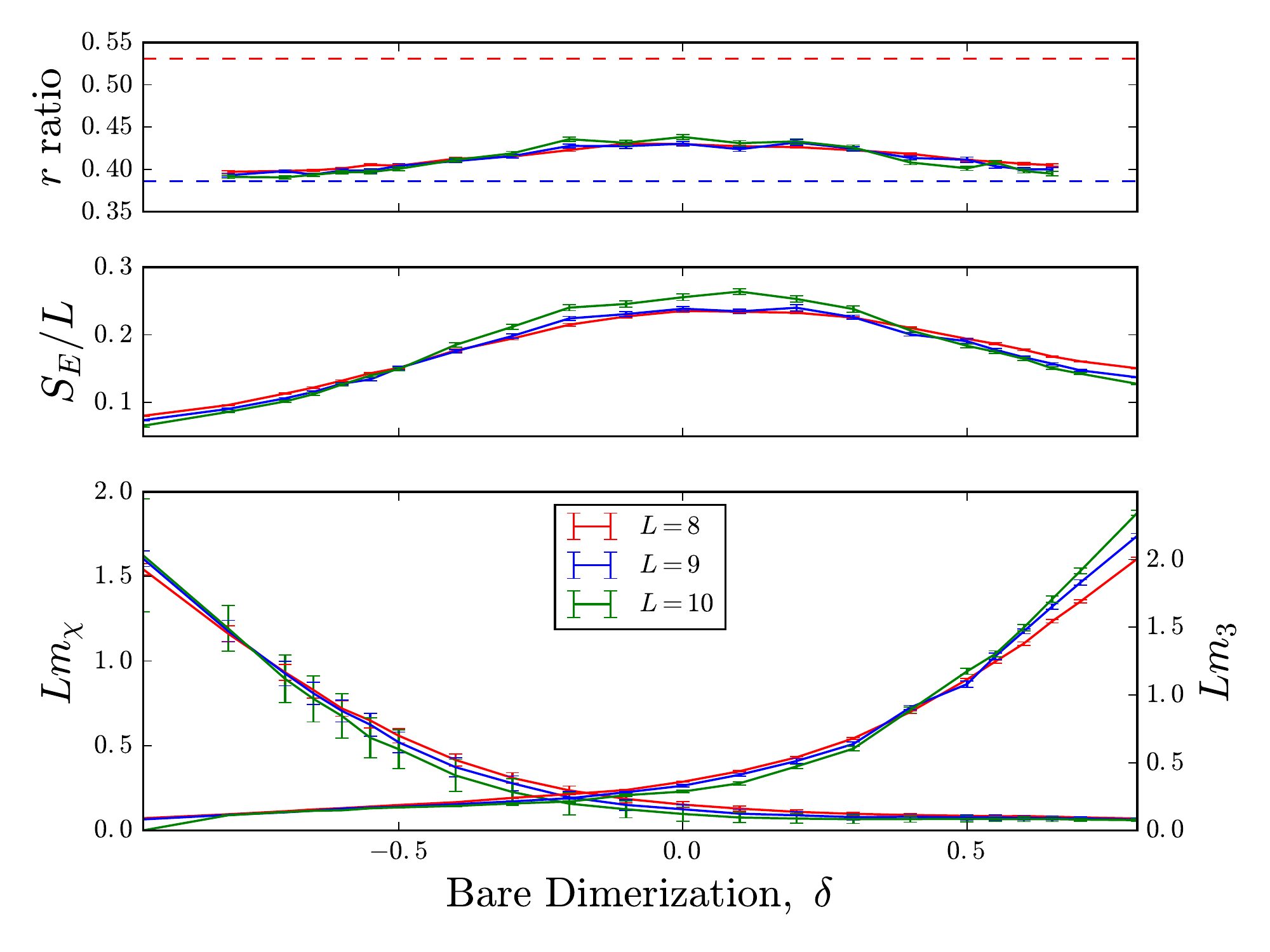}
\label{fig:subfig6wk}}
\end{figure}

\begin{figure}[t!]
\caption{$W=1.0$ for comparison}{
\includegraphics[width=0.95\columnwidth]{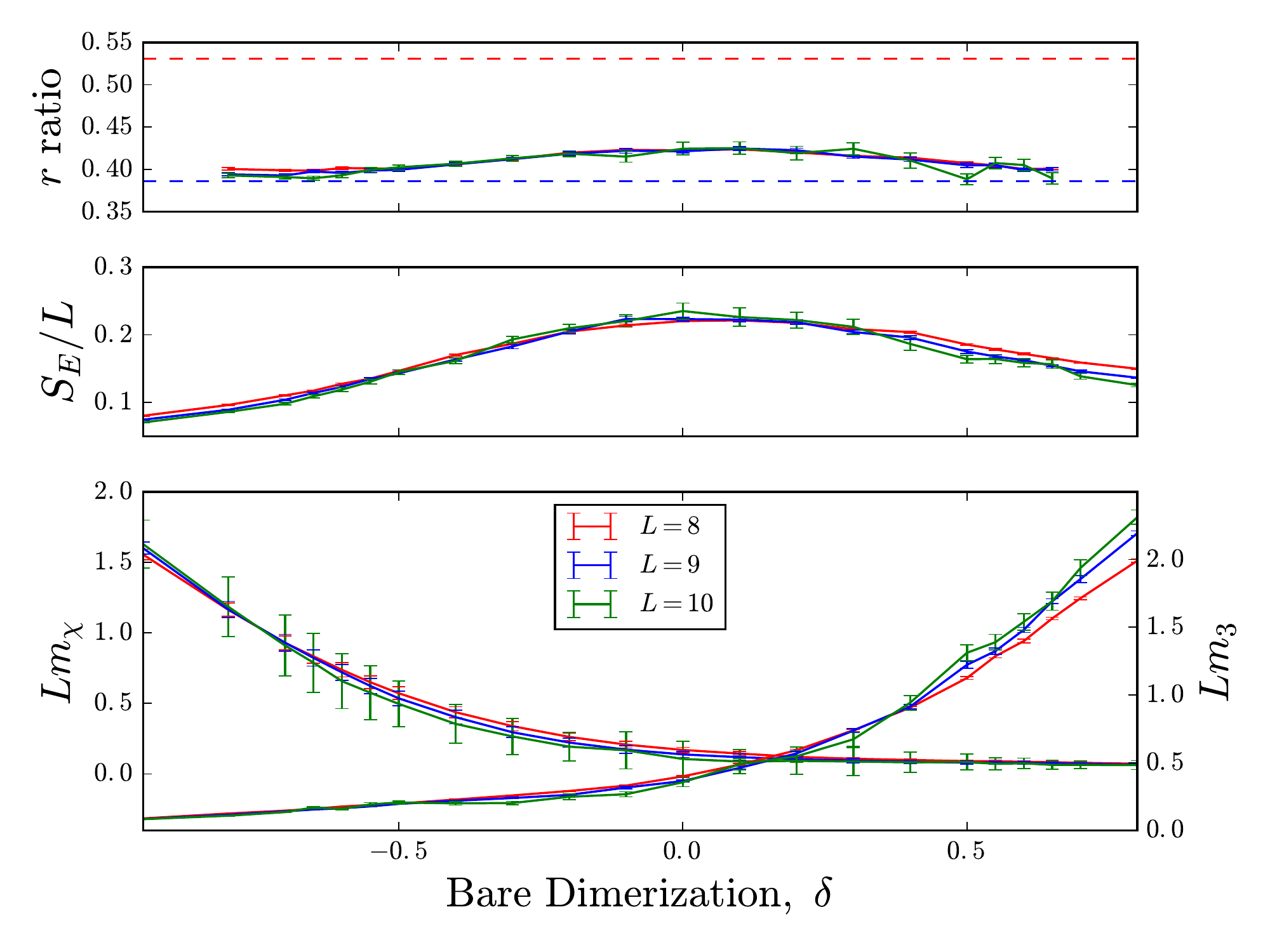}
\label{fig:subfig7wk}}
\end{figure}

Fig.~\ref{fig:subfig2wk}-\ref{fig:subfig7wk} present additional  data showing transitions in the level statistics, entanglement, and crossings in the $L$-scaled order parameters for weak disorder, used to determine the ETH phase boundary in Fig.~3 of the main text.

%\subfloat[][$W=0.025$]{
%\includegraphics[width=0.4\textwidth]{S3wk_W025.pdf}
%\label{fig:subfig1wk}}
%\qquad

\subsection{Strong Disorder}
\begin{figure}[t!]
\caption{`Strong Disorder' plot for $W=1.0$}{
\includegraphics[width=0.95\columnwidth]{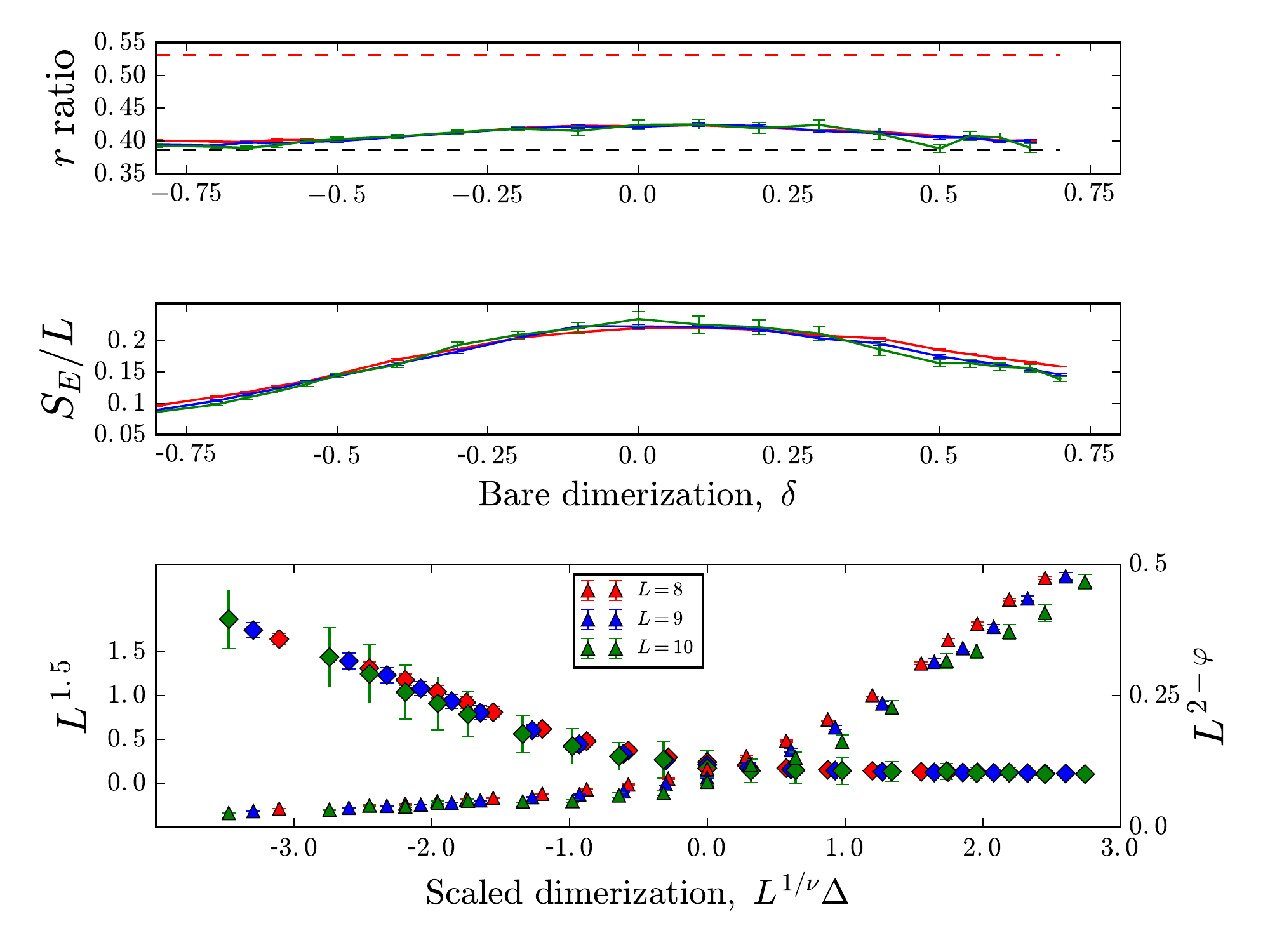}
\label{fig:subfig1sg}}
\end{figure}

\begin{figure}[t!]
\caption{Additional results for $W=1.5$}{
\includegraphics[width=0.95\columnwidth]{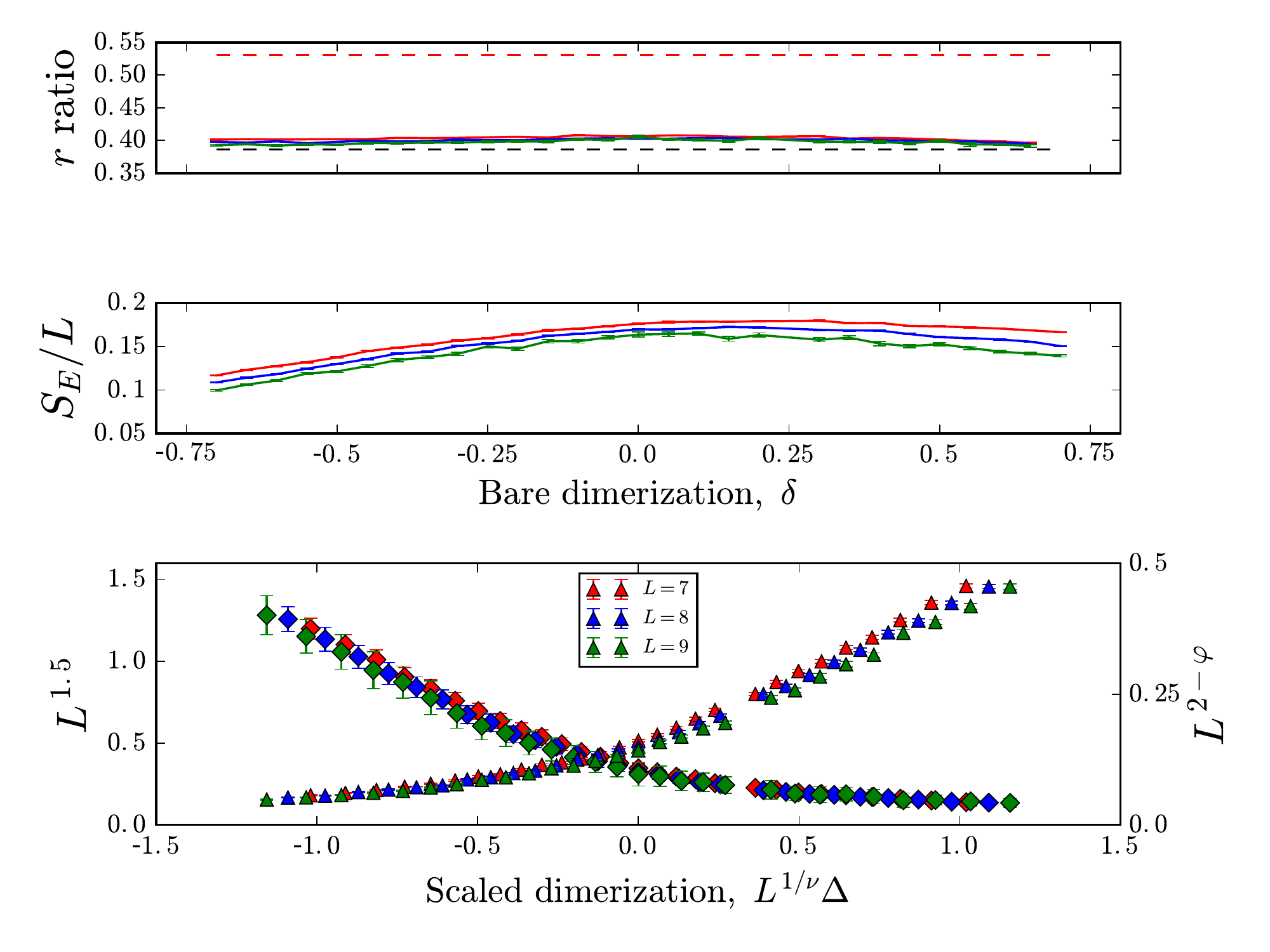}
\label{fig:subfig2sg}}
\end{figure}

\begin{figure}[H]
\caption{$W=2.0$ for comparison}{
\includegraphics[width=0.95\columnwidth]{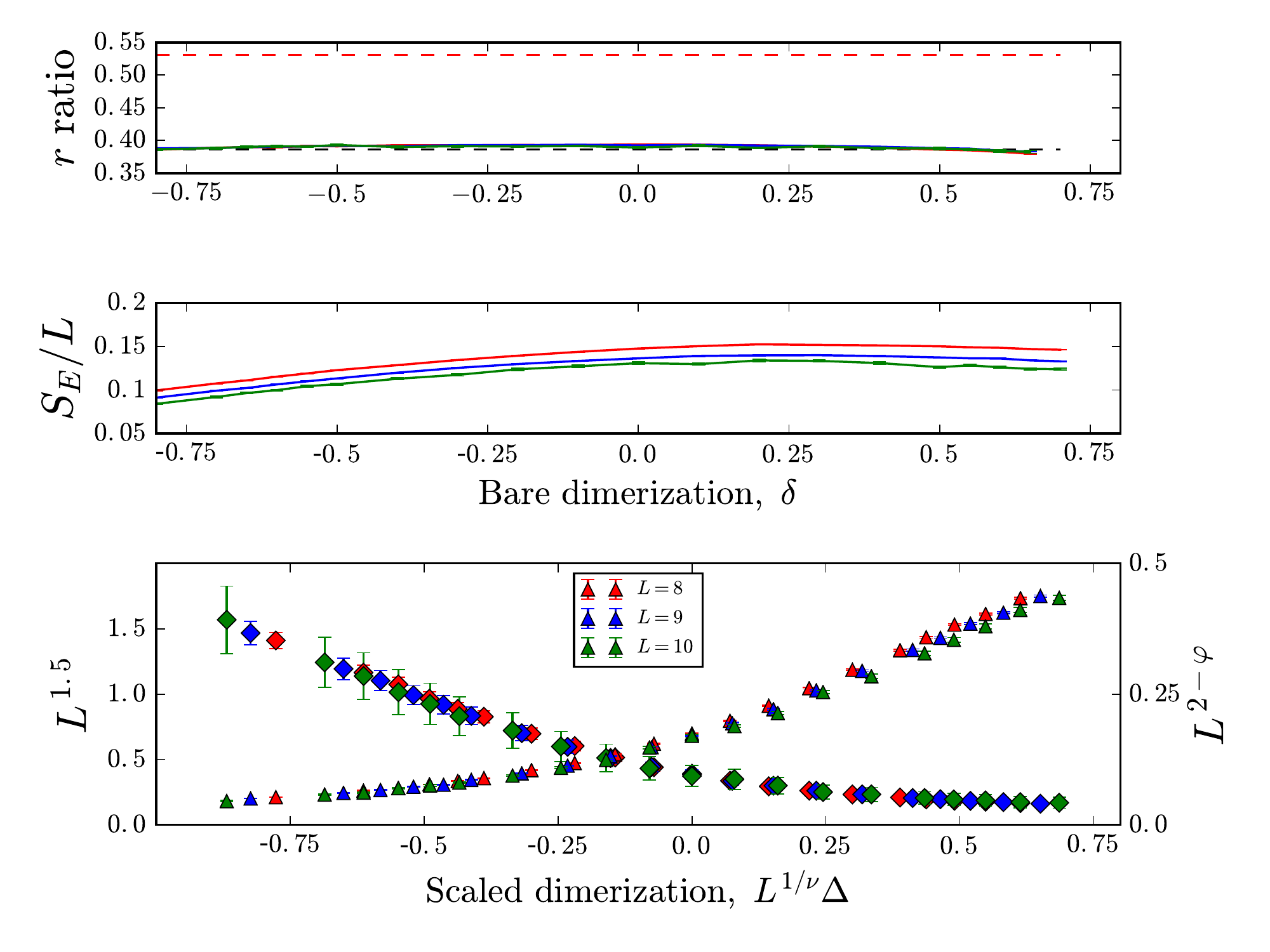}
\label{fig:subfig3sg}}
\end{figure}

\begin{figure}[H]
\caption{Additional results for $W=2.5$}{
\includegraphics[width=0.95\columnwidth]{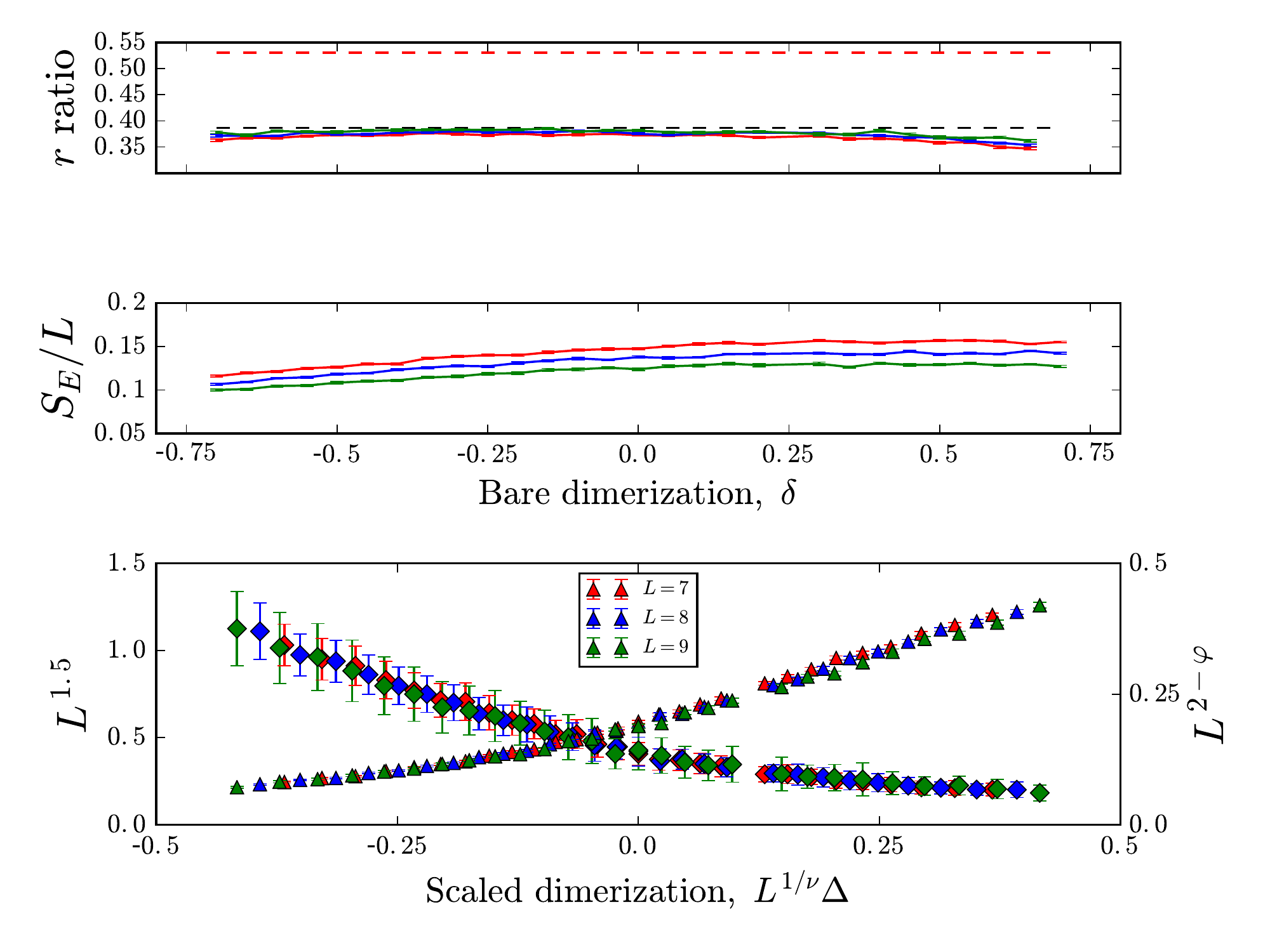}
\label{fig:subfig4sg}}
\end{figure}

Fig.~\ref{fig:subfig1sg}-\ref{fig:subfig4sg} present additional data showing Poisson level statistics and area law entanglement as one moves toward strong disorder ($W \sim 2$), as well as finite-size scaling collapse predicated on a direct infinite-randomness transition.  Note the improvement in the quality of collapse as we move to stronger disorder.

\end{appendix}
\bibliography{Potts_prb-latest.bib}

\end{document}